\documentclass[iop,appendixfloats,numberedappendix]{emulateapj}

\usepackage[caption=false]{subfig}
\usepackage{graphics}
\usepackage{graphicx}
\usepackage{rotating}
\usepackage{setspace}
\usepackage{gensymb}
\usepackage{textcomp}

\newcommand{\kms}{\,\mbox{km s$^{-1}$}}

\newcommand{\OIII}{[{O}{III}]}

\newcommand{\NII}{[{N}{II}]}

\newcommand{\Ha}{H$\alpha\,$}
\newcommand{\Hb}{H$\beta\,$}

\newcommand{\HII}{{H}{II}}

\shorttitle{}
\shortauthors{Rich et al.}

\begin{document}



\title{Composite Spectra in Merging U/LIRGs Caused by Shocks}

\author{J. A. Rich \altaffilmark{1,2},  L. J. Kewley\altaffilmark{3} \&  M. A. Dopita\altaffilmark{3,4}}
\email{jrich@obs.carnegiescience.edu.edu}
\altaffiltext{1}{Observatories of the Carnegie Institution of Washington, 813 Santa Barbara St., Pasadena, CA 91101}
\altaffiltext{2}{Spitzer Science Center, California Institute of Technology, 1200 E. California Blvd., Pasadena, CA 91125}
\altaffiltext{3}{Research School of Astronomy and Astrophysics, Australian National University, Cotter Rd., Weston ACT 2611, Australia}
\altaffiltext{4}{Astronomy Department, Faculty of Science, King Abdulaziz University, PO Box 80203, Jeddah, Saudi Arabia}

\date{\today}

\begin{abstract}
We present a key result from our optical integral field spectroscopic survey of 27 nearby ultra luminous and luminous infrared galaxies (U/LIRGs) from the Great Observatory All-sky LIRG Survey.  Using spatially resolved multi-component emission line fitting to trace the emission line ratios and velocity dispersion of the ionized gas, we quantify for the first time the widespread shock ionization in gas-rich merging U/LIRGs. Our results show a fractional contribution to the total observed \Ha flux from radiative shocks increasing from a few percent during early merger stages to upwards of 60\% of the observed optical emission line flux in late stage mergers. We compare our resolved spectroscopy to nuclear spectra and find that 3/4 of the galaxies in our sample that would be classified as ``composite'' based on optical spectroscopy are primarily characterized by a combination of star formation and merger-driven shocks. Our results have important implications for the interpretation of ``composite'' rest-frame optical spectra of U/LIRGs as starburst+AGN, as the shock emission combined with star formation can mimic ``composite'' optical spectra in the absence of any contribution from an AGN. 
\end{abstract}

\keywords{galaxies: evolution Ð galaxies: ISM Ð shock waves}

\section{Introduction}
Optical emission lines play a vital role in constraining the physics of the power sources in distant galaxies, typically dominated either by star forming \HII-regions or AGN. Early investigations investigated the utility of various emission lines in distinguishing between sources of ionizing radiation, including stars, AGN, shock fronts and planetary nebulae \citep{BPT81,VO87}. As larger and more refined spectroscopic samples became available and as radiative transfer models improved, the boundary lines separating star formation from AGN were solidified. \citet{Kauffmann03} and \citet{Kewley01a,Kewley06} established empirical and theoretical upper limits to the line ratios that can possibly be achieved in \HII~regions. The \NII/\Ha vs. \OIII/\Hb diagnostic is most commonly used in large studies, and has a region lying between pure star forming and pure AGN emission that is typically used to classify ``composite'' \HII~region + AGN spectra in galaxies. In some cases, galaxies may migrate from the empirical sequence of pure \HII~region emission towards the AGN portion of the diagnostic diagram as an AGN ignites and thus must have an increasing fractional contribution from a harder source of ionizing radiation \citep{Yuan10}.

In strong emission line galaxies, this ``composite'' emission is generally interpreted as a combination of AGN and \HII region emission (e.g. \citealt{Yuan10,Ellison11}).  In Starburst+LINER systems, however, the nature of observed composite emission may be a result of non-AGN sources. In particularly energetic systems, extended LINER emission has been observed due to starburst driven winds and merger-driven shocks \citep{Sharp10,Monreal10,Soto12}.  In \citet{Rich10,Rich11}, we presented results that revealed a mixture of LINER-like excitation driven by shocks and \HII-region emission due to ongoing intense star formation in three merging Ultraluminous and Luminous Infrared Galaxy (U/LIRG) systems. In a follow-up supplement we will present a detailed and expanded analysis of the 27 systems from our integral field spectroscopic (IFS) survey of nearby U/LIRGs.

In our survey we employ a combination of emission line ratio maps and emission line velocity dispersion derived from multi-component fits to separate shocked line emission from star formation. We also use a combination of optical, infrared and x-ray data to identify galaxies with a clear, strong AGN contribution. We analyze our IFS data with multi-component emission line fits and separate velocity components consistent with star formation from those consistent with shocks. Our results indicate an overall increase in total emission due to shocked gas as galaxy mergers progress in U/LIRGs, exceeding half of the total observed \Ha emission in some late-stage mergers. Moderate velocity resolution and spatial information are key factors in our analysis, though they come at the cost of sample size. 

Many large spectroscopic surveys, however, rely on single aperture spectra to determine galaxy properties (e.g. \citealt{Tremonti04,Ellison08,Stasinska08}). For nearby galaxies this typically means collecting only light from nuclear regions while at larger distances the observed spectrum can represent a large fraction of a galaxyÕs total emission. Although a loss of spatial information has thus far been necessary to survey statistically large samples of galaxies, some possible effects remain to be qualified. To this end, we compared the results of our spatially resolved analysis with single aperture nuclear spectra extracted from our IFS data. 

In this letter we present the results of the single-aperture analysis of our IFU survey of nearby U/LIRGs. \HII, ``composite'' and AGN classifications are derived from emission line ratio diagnostics using the nuclear spectra. We explain briefly the methods used to separate shock emission from star formation in our data cubes and compare nuclear spectra to our data cubes as well as multi-wavelength data useful for diagnosing AGN activity. We discuss a possible method for identifying the AGN in our sample using single-aperture spectra in the absence of IFS data.

\section{Observations and Analysis}
\begin{figure}
\centering
{\includegraphics[width=0.45\textwidth]{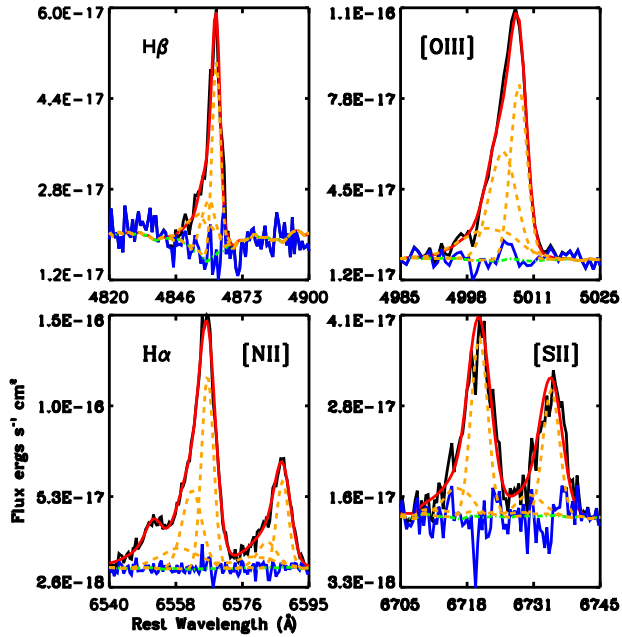}}
\caption{Example of a 3-component fit to a spectrum from a single spaxel in IRAS F23128-5919. Several strong emission lines are shown. the Total fit is plotted in red over the data in black, individual emission components are orange dashed lines, the continuum fit is a green dashed line and the residual fit between the total fit and the emission fit is shown in blue. This particular spaxel lies ${\sim}3$kpc SE of the northern nucleus in a region with multiple velocity components.}
\end{figure}

The Sample observed and analyzed for this letter is described in detail in \citet{Rich12_thesis} and a companion supplement in preparation. In short, our sample is composed of 27 systems taken from the Great Observatory All-Sky LIRG Survey (GOALS), which is itself a subset of the IR-selected Revised Bright Galaxy Sample \citep{Sanders03,Armus09}. These objects represent some of the most active and luminous systems in the nearby universe (z~$ <0.05$), including 4 ULIRGs. Most of the systems in our sample consist of two or more merging galaxies, we separate the systems into four isolated (``iso''), four widely separated systems (``a''), ten closely interacting mergers (``b''), and nine late stage, coalesced mergers (``cde'') as defined by \citet{Yuan10} and \citet{Veilleux02}. Separate from these merger stage classifications we also identify four galaxies which are clearly identifed as AGN in the far-infrared and x-ray \citep{Farrah07,Iwasawa09,Iwasawa11,Petric11}. These AGN are composed of one of the isolated systems and three of the closely interacting systems. The remaining systems classified in the merger sequence show weak or ambiguous indication of an AGN across multi-wavelength data.

\subsection{Integral Field Spectra}
Our observations were made using the Wide Field Spectrograph (WiFeS), an image-slicing, dual beam spectrograph on the Australian National University 2.3m telescope at Siding Spring Observatory \citep{Dopita07,Dopita10}. A single pointing consists of a blue and red cube with a FOV of 25\arcsec x38\arcsec sampled with 1\arcsec square spaxels. The blue cubes cover 3700-5500\AA~at R~${\sim}3000$ while the red cubes cover 5300-7000\AA~at a moderately high R~${\sim}7000$. The data were reduced using IRAF packages created for the WiFeS instrument, as described in detail in \citep{Rich10,Rich11,Rich12}. The resulting blue and red data cubes analyzed are flux-calibrated, sky-subtracted and where necessary mosaiced to cover a larger field of view.

We also analyze nuclear spectra from each system. For each data cube, a fixed 1 kpc circular aperture was extracted from the brightest optical nuclear region, combining several spaxels, ranging from 4 for the most distant systems to 32 for the nearest system, into a single spectrum. Each spaxel's spectrum was weighted based on the percentage of that spaxel that fell within the circular aperture, in the same fashion as \citet{Rich11}. 

\subsection{Data Analysis}
Our analysis focuses on the properties of the emission line gas in our sample. For each spectrum in a data cube, we first fit a stellar continuum with emission masked using a linear combination of stellar templates, with stellar population synthesis models from \citet{Gonzalez05} and fitting software which employs least-squares fitting derived from IBACKFIT and MPFIT \citep{Moustakas06,Markwardt09}. This continuum is subtracted and our own multi-component emission line fitting software is run on the remaining continuum-subtracted spectrum.

To determine the number of emission line components to fit, the \NII+\Ha region from each spectrum is fit with one, two and three gaussian components. Each fit is checked by eye and compared with neighboring fits and one two or three components are assigned depending on both a subjective goodness of fit as well as a comparison with the increase in a statistical measure of the goodness of fit. The width and redshift of each emission line component is then fixed and all strong emission lines are fit with the derived gaussians. An example of a three component fit from one of our systems is shown in Figure 1. Only emission line components with a peak flux with a signal-to-noise ratio greater than 3$\sigma$ were considered in our analysis. 

The nuclear spectra are fit in a similar fashion and corresponding emission line fluxes and widths are then extracted from the resulting fit. This is similar to the SDSS-fiber simulations in \citet{Rich11}, but no binning is performed in the spectral direction, maintaining the native spectral resolution.

\section{Resolved vs. Single Aperture Spectroscopy}
\begin{figure*}[htpb!]
\centering
{\includegraphics[width=\textwidth]{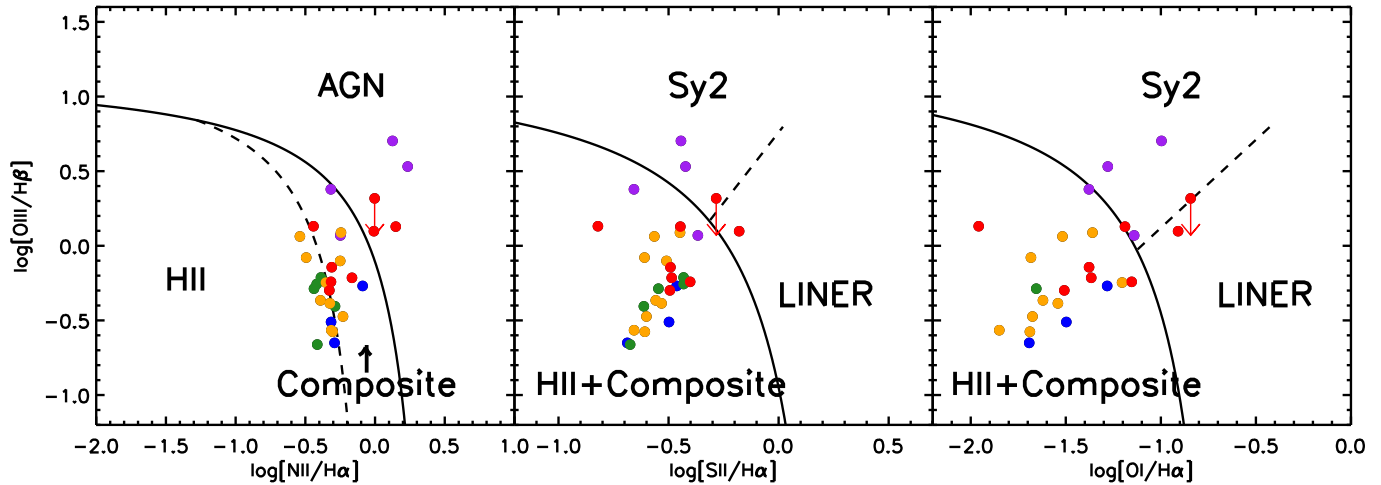}}
\caption{Line ratio diagnostic diagrams for the 1 kpc nuclear emission in the WiFeS GOALS sample. Blue, green, orange, red and purple points correspond to the isolated, a, b, cde and AGN classifications respectively. The nuclear spectrum for IRAS F12592+0436 is plotted with the measured upper limit for \OIII/\Hb.}
\end{figure*}

Our resolved spectroscopy reveals spectra that lie in the composite region of the diagnostic diagrams which have no apparent contribution from an AGN. Instead, the composite activity results from an increasing contribution to the optical spectra from shocks in late-stage mergers. In order to investigate whether these galaxies could be mistaken for composite \HII+AGN, we compare our extracted 1 kpc nuclear spectra to the systems' overall characteristics. 

Figure 2 shows diagnostic diagrams with line ratios calculated from the 1 kpc nuclear spectra for each system, color coded with respect to the galaxies merger class. In total, 12 of the systems in our sample lie in the ``composite'' region when a 1 kpc aperture is used to determine the galaxy type. The blue point in the composite region corresponds to the previously classified LINER nucleus in IRAS F18341-5732. Two of the AGN fall in the upper portion of the composite region as well. Figure 2 also shows that several middle and late-stage mergers with no clear indication of AGN activity lie in the ``composite'' region in both the nuclear spectra. In all, 9 of the 12 ``composite'' systems exhibit spectra that appear to be the result of a mixture of merger-driven shocks and \HII-region emission, rather than strictly a mix of AGN and \HII-regions. 

\section{Velocity Dispersion, Shocks and AGN}
We quantify the relative amount of shocks present in a galaxy using the spaxel-by-spaxel velocity dispersion, $\sigma$, as a proxy. Figure 3 shows the velocity dispersion distribution of the emission line fits. Lower-$\sigma$ values correspond to \HII-region emission and turbulent star formation, up until approximately $\sigma=90\kms$ where shocks dominate (e.g. \citealt{Genzel08,Green10}.  Clearly there is a shift in the $\sigma$-distribution towards more turbulent star formation and shocks as the mergers progress. 

One difference between the clearly identified AGN and the non-AGN in our sample lies in the emission line velocity dispersion distribution. Figure 3 shows the velocity dispersion distribution of the emission line fits in the different merger stages and the AGN in our sample. The shape of the non-AGN dominated systems is discussed in the companion supplement, but is dominated primarily by low-$\sigma$ star forming regions and higher-$\sigma$ (${\sim}100-200$ \kms) shocks, which become an increasingly dominant component of the optical emission at later merger stages. The AGN in Figure 3 show a small but significant contribution from emission line components with velocity dispersions well in excess of 250 \kms, significantly beyond the upper end of the velocity dispersions exhibited by low-velocity shocks in the non-AGN mergers.

\begin{figure}[htpb!]
\centering
{\includegraphics[width=0.40\textwidth]{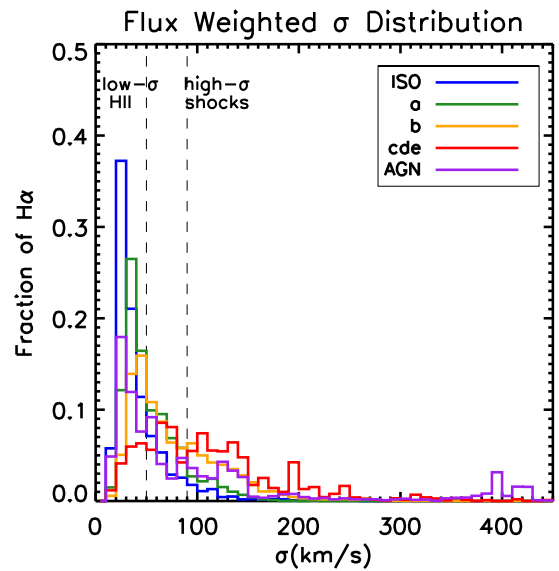}}
\caption{Distributions of emission line velocity dispersions as a function of merger class for our sample of 27 systems. The fractional distribution is flux-weighted to avoid over-weighting of low surface brightness regions. The strong AGN in our sample do not show as large a contribution between 90 and 200 km/s, but do show a bump in excess of 400 km/s.}
\end{figure}

The total contribution from shocks at each merger stage can be measured using the values shown in Figure 3. We separate the flux-weighted emission line velocity dispersion components into four separate velocity dispersion groups to investigate the relative contribution from star formation, shocks and AGN. We associate low-$\sigma$ emission with \HII-regions, mid-$\sigma$ with turbulent star forming regions, high-$\sigma$ with shocks and very high-$\sigma$ with AGN, with relative contribution from each shown in Figure 4.
 
The non-AGN systems exhibit nearly zero emission above a $\sigma$ cutoff of 350\kms, remaining dominated by star formation at early merger stages and shocks at the latest merger stages. It is worth noting that in the non-AGN, the emission is split nearly evenly between the turbulent star forming and shock-like component (55\% vs. 42\%, with the remainder in the low-$\sigma$ bin). This is likely a combination of beam smearing disk rotation and outflow: to properly disentangle these components in the unresolved nuclear regions of shock-dominated galaxies will require higher spatial resolution IFS.

The AGN, however, show a contribution to the total \Ha emission in excess of 10\% above the highest $\sigma$ cutoff. The majority of this very high $\sigma$ emission arises in regions coincident with the AGN in each system. When the 1 kpc apertures are considered, the fraction of emission with $\sigma>350\kms$ remains at nearly zero for the non-AGN and increases to 35\% for the AGN. This highest-$\sigma$ component is likely driven by the seyfert nuclei in the four AGN and may point to a method for discriminating between AGN and non-AGN in unresolved optical spectra given a high enough spectroscopic resolution, though further observations are necessary given our small sample size. AGN themselves may drive shocks, especially at the higher velocities consistent with the very high-$\sigma$ component: such velocities indeed require a non-stellar source (e.g. \citealt{Westmoquette12}).

\begin{figure}
\centering
{\includegraphics[width=0.45\textwidth]{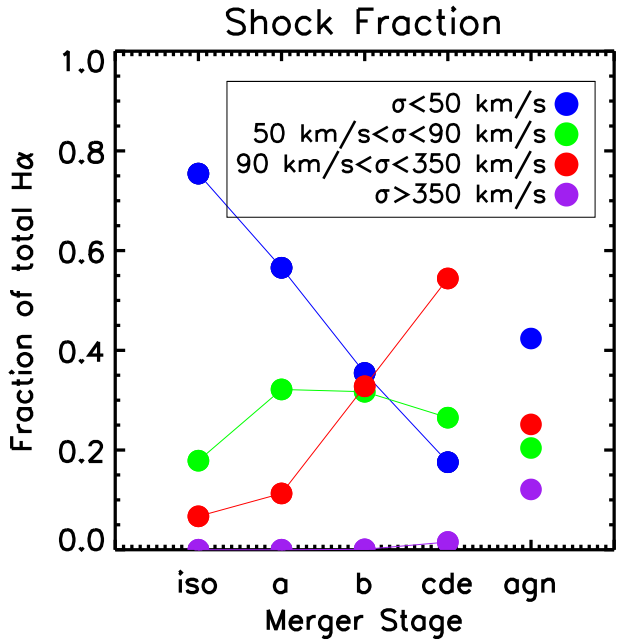}}
\caption{The fractional contribution in our resolved spectroscopic data of \HII-region ($\sigma<$50\kms), turbulent ($50\kms<\sigma<$90\kms), low velocity shocks (90\kms$<\sigma<$350\kms) and higher sigma components. In each bin, the total fraction of observed \Ha flux within each velocity dispersion range is calculated. The clearly defined AGN are the only systems with an appreciable contribution from components with $\sigma>$350\kms.}
\end{figure}

\section{discussion}
Clearly some composite galaxies are a combination of star formation and shock excitation, not simply starburst-AGN composites as suggested by their location on the starburst-AGN mixing sequence. Without spatially resolved spectroscopy with moderate velocity resolution it is difficult to determine whether the composite behavior in a nuclear spectrum is in fact the result of \HII+AGN or \HII+shock excitation, as is the case in some of our galaxies. This may affect the results of, for instance, \citet{Yuan10} where composite galaxies in late-stage mergers are considered a transitional class between pure \HII~emission and pure AGN emission in end-stage merger ULIRGs. There is likely less contamination of the composite region by shocks in non-interacting or more quiescent star-forming systems as the drivers of shock excitation, strong galactic winds and tidally induced gas flows, contribute far less energy to the ISM. To briefly summarize our findings:

\begin{itemize}

\item Using kpc aperture nuclear spectra, our IFS sample yields  12 ``composite'' systems. 3/4 of these composite systems are a mixture of \HII-region and shocked gas emission rather than purely AGN+starburst composite spectra

\item The systems clearly harboring AGN as evidenced by multi-wavelength data show a strong flux-weighted velocity dispersion component in our resolved spectra at relatively high $\sigma$ when compared to systems with shocks but no clear AGN. This component is primarily restricted to the nuclear regions and is in excess of 350\kms.

\end{itemize}

Without resolved spectroscopy, this mix of star formation and shocks can lead to a composite spectrum, depending on the location of the shocks, location of the aperture used and depending on the total fractional contribution to the optical emission from shocks. Indeed, the fact that shock-like line ratios are present outside of the star forming cores in many U/LIRGs (e.g. \citealt{Monreal10,Soto12}) implies that an admixture of \HII-region emission due to a nuclear starburst could result in a composite spectrum.  This mixture of shocks and star formation has also been observed in resolved spectroscopy of high redshift systems, with up to 25\% of line emission in some systems possibly emerging from shocks \citep{Newman12,Newman13}.

In many cases, a mixture of star formation, shock excitation, AGN activity and even old stellar populations may be the cause of composite emission, although even with resolved optical spectroscopy the contribution of each power source may remain ambiguous (e.g. \citealt{CidFernandes09,CidFernandes10,Alatalo11,Newman13}). The situation is further complicated by beam smearing in resolved spectroscopy and and other effects caused by low spatial resolution \citep{Yuan13}. 

Higher spatial and spectral resolution IFS observations analyzed using a combination of multiple optical emission lines and new theoretical models may allow observers to not only determine the fractional contribution from stars, shocks and AGN. In some systems, however, further multi-wavelength data may be necessary to disentangle the various power sources at work in many systems both in the local universe and at high redshift.

\begin{acknowledgements}
We thank the referee for their careful reading of the paper and insightful comments which improved the manuscript. The authors acknowledge ARC support under Discovery  project DP0984657. This research has made use of the NASA/IPAC Extragalactic Database (NED) which is operated by  the Jet Propulsion Laboratory, California Institute of Technology, under contract with the National  Aeronautics and Space Administration. This research has also made use of NASA's Astrophysics Data System, and of SAOImage DS9 \citep{joye03}, developed by the Smithsonian Astrophysical Observatory. 
\end{acknowledgements}

\bibliographystyle{apj}

\begin{thebibliography}{36}
\expandafter\ifx\csname natexlab\endcsname\relax\def\natexlab#1{#1}\fi

\bibitem[{{Alatalo} {et~al.}(2011){Alatalo}, {Blitz}, {Young}, {Davis},
  {Bureau}, {Lopez}, {Cappellari}, {Scott}, {Shapiro}, {Crocker},
  {Mart{\'{\i}}n}, {Bois}, {Bournaud}, {Davies}, {de Zeeuw}, {Duc}, {Emsellem},
  {Falc{\'o}n-Barroso}, {Khochfar}, {Krajnovi{\'c}}, {Kuntschner}, {Lablanche},
  {McDermid}, {Morganti}, {Naab}, {Oosterloo}, {Sarzi}, {Serra}, \&
  {Weijmans}}]{Alatalo11}
{Alatalo}, K., {et~al.} 2011, \apj, 735, 88

\bibitem[{{Armus} {et~al.}(2009){Armus}, {Mazzarella}, {Evans}, {Surace},
  {Sanders}, {Iwasawa}, {Frayer}, {Howell}, {Chan}, {Petric}, {Vavilkin},
  {Kim}, {Haan}, {Inami}, {Murphy}, {Appleton}, {Barnes}, {Bothun}, {Bridge},
  {Charmandaris}, {Jensen}, {Kewley}, {Lord}, {Madore}, {Marshall},
  {Melbourne}, {Rich}, {Satyapal}, {Schulz}, {Spoon}, {Sturm}, {U}, {Veilleux},
  \& {Xu}}]{Armus09}
{Armus}, L., {et~al.} 2009, \pasp, 121, 559

\bibitem[{{Baldwin} {et~al.}(1981){Baldwin}, {Phillips}, \&
  {Terlevich}}]{BPT81}
{Baldwin}, J.~A., {Phillips}, M.~M., \& {Terlevich}, R. 1981, \pasp, 93, 5

\bibitem[{{Cid Fernandes} {et~al.}(2009){Cid Fernandes}, {Schlickmann},
  {Stasinska}, {Asari}, {Gomes}, {Schoenell}, {Mateus}, \&
  {Sodr{\'e}}}]{CidFernandes09}
{Cid Fernandes}, R., {Schlickmann}, M., {Stasinska}, G., {Asari}, N.~V.,
  {Gomes}, J.~M., {Schoenell}, W., {Mateus}, A., \& {Sodr{\'e}}, Jr., L. 2009,
  in Astronomical Society of the Pacific Conference Series, Vol. 408,
  Astronomical Society of the Pacific Conference Series, ed. {W.~Wang, Z.~Yang,
  Z.~Luo, \& Z.~Chen}, 122--+

\bibitem[{{Cid Fernandes} {et~al.}(2010){Cid Fernandes}, {Stasi{\'n}ska},
  {Schlickmann}, {Mateus}, {Vale Asari}, {Schoenell}, \&
  {Sodr{\'e}}}]{CidFernandes10}
{Cid Fernandes}, R., {Stasi{\'n}ska}, G., {Schlickmann}, M.~S., {Mateus}, A.,
  {Vale Asari}, N., {Schoenell}, W., \& {Sodr{\'e}}, L. 2010, \mnras, 403, 1036

\bibitem[{{Dopita} {et~al.}(2007){Dopita}, {Hart}, {McGregor}, {Oates},
  {Bloxham}, \& {Jones}}]{Dopita07}
{Dopita}, M., {Hart}, J., {McGregor}, P., {Oates}, P., {Bloxham}, G., \&
  {Jones}, D. 2007, \apss, 310, 255

\bibitem[{{Dopita} {et~al.}(2010){Dopita}, {Rhee}, {Farage}, {McGregor},
  {Bloxham}, {Green}, {Roberts}, {Neilson}, {Wilson}, {Young}, {Firth},
  {Busarello}, \& {Merluzzi}}]{Dopita10}
{Dopita}, M., {et~al.} 2010, \apss, 95

\bibitem[{{Ellison} {et~al.}(2011){Ellison}, {Patton}, {Mendel}, \&
  {Scudder}}]{Ellison11}
{Ellison}, S.~L., {Patton}, D.~R., {Mendel}, J.~T., \& {Scudder}, J.~M. 2011,
  \mnras, 1541

\bibitem[{{Ellison} {et~al.}(2008){Ellison}, {Patton}, {Simard}, \&
  {McConnachie}}]{Ellison08}
{Ellison}, S.~L., {Patton}, D.~R., {Simard}, L., \& {McConnachie}, A.~W. 2008,
  \aj, 135, 1877

\bibitem[{{Farrah} {et~al.}(2007){Farrah}, {Bernard-Salas}, {Spoon}, {Soifer},
  {Armus}, {Brandl}, {Charmandaris}, {Desai}, {Higdon}, {Devost}, \&
  {Houck}}]{Farrah07}
{Farrah}, D., {et~al.} 2007, \apj, 667, 149

\bibitem[{{Gonz{\'a}lez Delgado} {et~al.}(2005){Gonz{\'a}lez Delgado},
  {Cervi{\~n}o}, {Martins}, {Leitherer}, \& {Hauschildt}}]{Gonzalez05}
{Gonz{\'a}lez Delgado}, R.~M., {Cervi{\~n}o}, M., {Martins}, L.~P.,
  {Leitherer}, C., \& {Hauschildt}, P.~H. 2005, \mnras, 357, 945

\bibitem[{{Iwasawa} {et~al.}(2009){Iwasawa}, {Sanders}, {Evans}, {Mazzarella},
  {Armus}, \& {Surace}}]{Iwasawa09}
{Iwasawa}, K., {Sanders}, D.~B., {Evans}, A.~S., {Mazzarella}, J.~M., {Armus},
  L., \& {Surace}, J.~A. 2009, \apjl, 695, L103

\bibitem[{{Iwasawa} {et~al.}(2011){Iwasawa}, {Sanders}, {Teng}, {U}, {Armus},
  {Evans}, {Howell}, {Komossa}, {Mazzarella}, {Petric}, {Surace}, {Vavilkin},
  {Veilleux}, \& {Trentham}}]{Iwasawa11}
{Iwasawa}, K., {et~al.} 2011, \aap, 529, A106

\bibitem[{{Joye} \& {Mandel}(2003)}]{joye03}
{Joye}, W.~A., \& {Mandel}, E. 2003, in Astronomical Society of the Pacific
  Conference Series, Vol. 295, Astronomical Data Analysis Software and Systems
  XII, ed. {H.~E.~Payne, R.~I.~Jedrzejewski, \& R.~N.~Hook}, 489--+

\bibitem[{{Kauffmann} {et~al.}(2003){Kauffmann}, {Heckman}, {Tremonti},
  {Brinchmann}, {Charlot}, {White}, {Ridgway}, {Brinkmann}, {Fukugita}, {Hall},
  {Ivezi{\'c}}, {Richards}, \& {Schneider}}]{Kauffmann03}
{Kauffmann}, G., {et~al.} 2003, \mnras, 346, 1055

\bibitem[{{Kewley} {et~al.}(2001){Kewley}, {Dopita}, {Sutherland}, {Heisler},
  \& {Trevena}}]{Kewley01a}
{Kewley}, L.~J., {Dopita}, M.~A., {Sutherland}, R.~S., {Heisler}, C.~A., \&
  {Trevena}, J. 2001, \apj, 556, 121

\bibitem[{{Kewley} {et~al.}(2006){Kewley}, {Groves}, {Kauffmann}, \&
  {Heckman}}]{Kewley06}
{Kewley}, L.~J., {Groves}, B., {Kauffmann}, G., \& {Heckman}, T. 2006, \mnras,
  372, 961

\bibitem[{{Markwardt}(2009)}]{Markwardt09}
{Markwardt}, C.~B. 2009, in Astronomical Society of the Pacific Conference
  Series, Vol. 411, Astronomical Society of the Pacific Conference Series, ed.
  {D.~A.~Bohlender, D.~Durand, \& P.~Dowler}, 251--+

\bibitem[{{Monreal-Ibero} {et~al.}(2010){Monreal-Ibero}, {Arribas}, {Colina},
  {Rodr{\'{\i}}guez-Zaur{\'{\i}}n}, {Alonso-Herrero}, \&
  {Garc{\'{\i}}a-Mar{\'{\i}}n}}]{Monreal10}
{Monreal-Ibero}, A., {Arribas}, S., {Colina}, L.,
  {Rodr{\'{\i}}guez-Zaur{\'{\i}}n}, J., {Alonso-Herrero}, A., \&
  {Garc{\'{\i}}a-Mar{\'{\i}}n}, M. 2010, \aap, 517, A28+

\bibitem[{{Moustakas} \& {Kennicutt}(2006)}]{Moustakas06}
{Moustakas}, J., \& {Kennicutt}, Jr., R.~C. 2006, \apjs, 164, 81

\bibitem[{{Newman} {et~al.}(2012){Newman}, {Genzel}, {F{\"o}rster-Schreiber},
  {Shapiro Griffin}, {Mancini}, {Lilly}, {Renzini}, {Bouch{\'e}}, {Burkert},
  {Buschkamp}, {Carollo}, {Cresci}, {Davies}, {Eisenhauer}, {Genel}, {Hicks},
  {Kurk}, {Lutz}, {Naab}, {Peng}, {Sternberg}, {Tacconi}, {Vergani}, {Wuyts},
  \& {Zamorani}}]{Newman12}
{Newman}, S.~F., {et~al.} 2012, \apj, 761, 43

\bibitem[{{Newman} {et~al.}(2013){Newman}, {Buschkamp}, {Genzel}, {Forster
  Schreiber}, {Kurk}, {Sternberg}, {Gnat}, {Rosario}, {Mancini}, {Lilly},
  {Renzini}, {Burkert}, {Carollo}, {Cresci}, {Davies}, {Eisenhauer}, {Genel},
  {Shapiro Griffin}, {Hicks}, {Lutz}, {Naab}, {Peng}, {Tacconi}, {Wuyts},
  {Zamorani}, {Vergani}, \& {Weiner}}]{Newman13}
---. 2013, ArXiv e-prints

\bibitem[{{Petric} {et~al.}(2011){Petric}, {Armus}, {Howell}, {Chan},
  {Mazzarella}, {Evans}, {Surace}, {Sanders}, {Appleton}, {Charmandaris},
  {D{\'{\i}}az-Santos}, {Frayer}, {Haan}, {Inami}, {Iwasawa}, {Kim}, {Madore},
  {Marshall}, {Spoon}, {Stierwalt}, {Sturm}, {U}, {Vavilkin}, \&
  {Veilleux}}]{Petric11}
{Petric}, A.~O., {et~al.} 2011, \apj, 730, 28

\bibitem[{{Rich} {et~al.}(2010){Rich}, {Dopita}, {Kewley}, \& {Rupke}}]{Rich10}
{Rich}, J.~A., {Dopita}, M.~A., {Kewley}, L.~J., \& {Rupke}, D.~S.~N. 2010,
  \apj, 721, 505

\bibitem[{{Rich} {et~al.}(2011){Rich}, {Kewley}, \& {Dopita}}]{Rich11}
{Rich}, J.~A., {Kewley}, L.~J., \& {Dopita}, M.~A. 2011, \apj, 734, 87

\bibitem[{{Rich} {et~al.}(2012){Rich}, {Torrey}, {Kewley}, {Dopita}, \&
  {Rupke}}]{Rich12}
{Rich}, J.~A., {Torrey}, P., {Kewley}, L.~J., {Dopita}, M.~A., \& {Rupke},
  D.~S.~N. 2012, \apj, 753, 5

\bibitem[{{Rich}(2012)}]{Rich12_thesis}
{Rich Jr}, J.~A.~S. 2012, PhD thesis, University of Hawai'i at Manoa

\bibitem[{{Sanders} {et~al.}(2003){Sanders}, {Mazzarella}, {Kim}, {Surace}, \&
  {Soifer}}]{Sanders03}
{Sanders}, D.~B., {Mazzarella}, J.~M., {Kim}, D., {Surace}, J.~A., \& {Soifer},
  B.~T. 2003, \aj, 126, 1607

\bibitem[{{Sharp} \& {Bland-Hawthorn}(2010)}]{Sharp10}
{Sharp}, R.~G., \& {Bland-Hawthorn}, J. 2010, \apj, 711, 818

\bibitem[{{Soto} \& {Martin}(2012)}]{Soto12}
{Soto}, K.~T., \& {Martin}, C.~L. 2012, \apjs, 203, 3

\bibitem[{{Stasi{\'n}ska} {et~al.}(2008){Stasi{\'n}ska}, {Vale Asari}, {Cid
  Fernandes}, {Gomes}, {Schlickmann}, {Mateus}, {Schoenell}, \&
  {Sodr{\'e}}}]{Stasinska08}
{Stasi{\'n}ska}, G., {Vale Asari}, N., {Cid Fernandes}, R., {Gomes}, J.~M.,
  {Schlickmann}, M., {Mateus}, A., {Schoenell}, W., \& {Sodr{\'e}}, Jr., L.
  2008, \mnras, 391, L29

\bibitem[{{Tremonti} {et~al.}(2004){Tremonti}, {Heckman}, {Kauffmann},
  {Brinchmann}, {Charlot}, {White}, {Seibert}, {Peng}, {Schlegel}, {Uomoto},
  {Fukugita}, \& {Brinkmann}}]{Tremonti04}
{Tremonti}, C.~A., {et~al.} 2004, \apj, 613, 898

\bibitem[{{Veilleux} \& {Osterbrock}(1987)}]{VO87}
{Veilleux}, S., \& {Osterbrock}, D.~E. 1987, \apjs, 63, 295

\bibitem[{{Veilleux} \& {Rupke}(2002)}]{Veilleux02}
{Veilleux}, S., \& {Rupke}, D.~S. 2002, \apjl, 565, L63

\bibitem[{{Westmoquette} {et~al.}(2012){Westmoquette}, {Clements}, {Bendo}, \&
  {Khan}}]{Westmoquette12}
{Westmoquette}, M.~S., {Clements}, D.~L., {Bendo}, G.~J., \& {Khan}, S.~A.
  2012, \mnras, 424, 416

\bibitem[{{Yuan} {et~al.}(2010){Yuan}, {Kewley}, \& {Sanders}}]{Yuan10}
{Yuan}, T., {Kewley}, L.~J., \& {Sanders}, D.~B. 2010, \apj, 709, 884

\bibitem[{{Yuan} {et~al.}(2013){Yuan}, {Kewley}, \& {Rich}}]{Yuan13}
{Yuan}, T.-T., {Kewley}, L.~J., \& {Rich}, J. 2013, \apj, 767, 106

\end{thebibliography}

\end{document}